\documentclass[10pt,conference]{IEEEtran}
\IEEEoverridecommandlockouts

\usepackage{cite}
\usepackage{amsmath,amssymb,amsfonts}
\usepackage{algorithmic}
\usepackage{graphicx}
\usepackage{textcomp}
\usepackage{xcolor}
\usepackage{svg}
\usepackage{todonotes}
\usepackage{longtable}
\usepackage{booktabs}
\usepackage{array}
\usepackage{tcolorbox}
\usepackage{tabularx}
\usepackage[table]{xcolor}
\usepackage{xspace}

\newif\ifdraft
\drafttrue 

    \def\BibTeX{{\rm B\kern-.05em{\sc i\kern-.025em b}\kern-.08em
        T\kern-.1667em\lower.7ex\hbox{E}\kern-.125emX}
    \fi
}

\begin{document}

\title{Beyond the Prompt: Linking What Developers Ask, Do, and Understand with Coding Agents}


\newif\ifanonymous
\anonymousfalse
\ifanonymous
\author{\IEEEauthorblockN{Anonymous Authors}
\IEEEauthorblockA{\textit{}} }
\else
\author{\IEEEauthorblockN{Yunhan Qiao\textsuperscript{1,*}\qquad Summit Haque\textsuperscript{1}\qquad Christopher Hundhausen\textsuperscript{1}}
\IEEEauthorblockA{\textsuperscript{1}Oregon State University, School of Electrical Engineering and Computer Science, Corvallis, OR, USA}
\thanks{\textsuperscript{*}Corresponding author.}
}
\fi


\maketitle

\begin{abstract}
Coding agents can now change code for developers, who describe goals, supply context, and respond to the agent's work. Yet prompts, screen activity, and task success each tell only part of this story. We present Say, Do, Understand, an end-to-end workflow for analyzing what developers write to an agent, what they do while it works, and what they can explain afterwards. The workflow has five stages (Capture, Prepare, Analyze, Integrate, and Interpret) and three instruments: a prompt codebook, a scheme for coding screen-recorded activities and events, and separate rubrics for explaining the process and the solution. We applied it in an observational study of ten experienced developers who used GitHub Copilot on an unfamiliar codebase. Crossing task performance with understanding produced four personas. The two measures agreed for eight developers but split for two: one passed most tests but could not explain the solution, and another passed few tests but explained it well. In this sample, the personas that most often asked the agent to check its work spent the least time testing on their own. These patterns are descriptive and do not generalize beyond the sample. We recommend the workflow to computing educators, industry practitioners, and researchers to adapt and evaluate human--AI communication in software engineering.
\end{abstract}


\section{Introduction}
\label{sec:introduction}

Coding agents let developers request changes to unfamiliar codebases in natural language. An agent can inspect files, plan changes, edit multiple components, execute commands, and run tests~\cite{yang2024swe,wang2025openhands}. Developers contribute by deciding what to ask, supplying relevant evidence, and responding to the work. These exchanges connect language with action. A prompt may identify a problem, but its meaning in the development process also depends on what the developer does and can explain afterwards. Adding features to an unfamiliar existing codebase through its existing backend API is a core software evolution task. As coding agents change how developers carry out such tasks, understanding how developers communicate with them becomes a software maintenance and evolution problem.

Understanding these exchanges matters to several audiences. Computing educators need ways to examine how students communicate with AI while assessing what students understand. Industry practitioners need evidence about how developers communicate with tools that modify existing systems. Researchers need instruments that connect communication with observable activity and independently assessed outcomes. In each setting, a working implementation matters, but it does not show whether a person understands the change. That distinction matters for later debugging, maintenance, review, and modification.

Existing research offers several ways to examine AI-assisted programming. Studies analyze how developers communicate requirements, revise prompts, control generated output, and validate suggestions~\cite{barke2023grounded,liang2024large,wang2024investigating}. Work in computing education also examines GenAI coding assistant usage and programming performance in existing codebases~\cite{kazemitabaar2024codeaid,qiao2026systematic,shihab2025effects}. These approaches provide complementary evidence. Prompt analysis describes what people ask for. Behavioral analysis describes visible activity. Task assessment describes what an implementation accomplishes. No single stream shows what a developer can explain about the process or the resulting solution. Recent work also calls for more study of how people communicate with coding agents~\cite{wang2026humans}.

What is missing is a practical way to connect these forms of evidence. Asking an agent to check its work is a communication act. Running a test is a visible action. Explaining why a change works shows understanding. Treating these as interchangeable hides the relationships we want to study. Likewise, time spent viewing code describes activity; it does not show why the developer read the code or whether they understood it. A combined account needs both links between the streams and clear limits on what each one means.

An explanation also needs a reference point. A general account of how a feature should work may not describe the code a participant actually produced. We therefore separate explanations of the development process from explanations of the submitted implementation. Assessing them separately gives credit for a coherent process account even when the implementation is incomplete.

We present \textit{Say, Do, Understand}, an end-to-end workflow, from collecting evidence to reporting patterns, for analyzing human--AI communication in programming. \textit{Say} captures what people write to the AI through prompts. \textit{Do} captures screen-recorded activity states and discrete events. \textit{Understand} captures what people explain afterwards about their process and implemented solution. The workflow has five stages: Capture, Prepare, Analyze, Integrate, and Interpret. We analyze the three streams in parallel, then combine their outputs in participant-level profiles. Interpretation identifies patterns and states the limits of the evidence.

We ask two research questions:
\begin{description}
    \item[RQ1:] How can prompts, behavior, and understanding be analyzed together in an end-to-end workflow?
    \item[RQ2:] What does applying the workflow reveal about how experienced developers communicate with a coding agent?
\end{description}

We address RQ1 by specifying the stages, instruments, and analysis principles. We address RQ2 through a worked application with ten experienced developers using GitHub Copilot on an unfamiliar MERN-stack application. Participants completed sequential frontend tasks during sessions lasting up to two hours and 15 minutes. We analyzed their prompts, screen recordings, test outcomes, and post-task explanations. Crossing performance with understanding gave four quadrants, which we describe as personas; the two measures agreed for eight participants but diverged for two: one passed most tests but could not explain the solution, and another passed few but explained it well. In this sample, the personas that most often asked the agent to check its work spent the least time testing on their own.

This paper makes three contributions:
\begin{itemize}
    \item An end-to-end workflow that connects prompt, behavioral, and explanation evidence while keeping clear what each stream can show.
    \item A set of analysis instruments (prompt codes, video states and events, and separate rubrics for process and solution understanding) with an LLM-assisted coding procedure that researchers refine until agreement is reliable.
    \item A worked application that summarizes participant-level profiles as four personas, showing patterns that any single stream would leave incomplete.
\end{itemize}

The contribution is the workflow and its illustrated use. In this application, we use Specify/Delegate/Verify to describe each persona; other analyses can use a different lens. Computing educators, industry practitioners, and researchers can adapt the workflow, but its reuse in these settings has not yet been tested. Our application supports descriptive, participant-level patterns, not temporal or causal accounts of communication.

\section{Related Work}
\label{sec:related}

\subsection{Interaction Frameworks and Human--AI Communication}

Human--AI communication involves coordinating work as well as exchanging instructions. Klein et al. describe joint human--agent activity through mutual predictability, directability, and common ground~\cite{klien2004ten}. Mixed-initiative interaction addresses who initiates action and how automated services stay compatible with human direction~\cite{horvitz1999principles}. Human--AI interaction guidelines turn related concerns into design recommendations, such as communicating capabilities and supporting correction~\cite{amershi2019guidelines}. These accounts help identify what an analysis of communication should consider. They do not, however, say how to combine prompts, visible activity, and explanations of implemented work.

Studies of coding assistants describe acceleration, where developers pursue a solution they already have in mind, and exploration, where they investigate unfamiliar approaches or work past an impasse~\cite{barke2023grounded}. We position Say, Do, Understand alongside interaction frameworks for studying programming with AI~\cite{mozannar2024reading,park2026delegation}. Our contribution is a workflow for processing evidence, through which researchers can apply such a framework. A framework can guide interpretation while the prompt codes, activities, and understanding measures stay open to separate inspection. In our application, Specify/Delegate/Verify serves as this optional lens during integration.

\subsection{Programming with GenAI}

Research documents difficulties in communicating requirements and controlling generated output~\cite{liang2024large}. Other studies examine expectations, trust, and validation of suggestions~\cite{wang2024investigating}, along with concerns about output accuracy and project context~\cite{sergeyuk2025using}. Coding agents extend the exchange beyond single completions. SWE-agent and OpenHands navigate repositories, edit files, execute commands, and respond to execution feedback~\cite{yang2024swe,wang2025openhands}. Communication can therefore span exploration, implementation, and debugging across several files.

Studies of professional practice also describe developers using their expertise to constrain agent behavior and keep control over design and implementation~\cite{huang2025professional}. These accounts motivate examining how requests relate to the actions around them. A prompt can ask for inspection or include an error message, but it cannot show whether the developer later checked the application. Conversely, a screen recording may show a test run without revealing what the developer understood afterwards. Our workflow keeps these observations separate before examining them together.

GenAI in software engineering education is another relevant setting~\cite{choudhuri2024howfar,kazemitabaar2024codeaid,qiao2026systematic}. Shihab et al. studied computing students completing brownfield tasks with GitHub Copilot, examining effectiveness, efficiency, and programming process~\cite{shihab2025effects}. We adapt their behavioral coding approach for experienced developers working with a coding agent. The educational value of our workflow is still prospective. Using the instruments with professionals does not show that they measure student understanding equally well.

\subsection{Comprehension and Metacognition}

Code-comprehension research uses many kinds of tasks and measures~\cite{wyrich202340}. Field studies describe comprehension activities in developers' daily work~\cite{xia2017measuring}, and experiments describe how experts approach unfamiliar code~\cite{koenemann1991expert}. Research on computational notebooks similarly examines how people make sense of unfamiliar artifacts~\cite{chattopadhyay2023make}. These approaches show how people engage with code. However, the time spent on an activity does not show whether a developer can accurately explain the implementation.

Evaluation methods include comprehension questions, modification tasks, and confidence ratings~\cite{wyrich202340}. Jask generates questions and answers about learners' own Java code~\cite{santos2022jask}. Eye tracking and physiological measures add evidence about attention, confusion, and code understanding~\cite{flint2026exploratory,peitek2022correlates,abdelsalam2026effect,saraiva2025no}. Our approach focuses on post-task explanations. It separates a coherent account of the development process from a correct explanation of the implemented solution.

This distinction also links the workflow to metacognitive accounts of GenAI use~\cite{tankelevitch2024metacognitive}. We assess what participants can explain about their approach, progress, decisions, and outcome. We separately assess their solution explanations against the code they committed. These measures do not observe cognition directly or show learning gains. They provide evidence of demonstrated understanding alongside task performance and behavior.

\subsection{Combining Evidence and Assisted Analysis}

Multi-source assessment is a relevant methodological precedent~\cite{hundhausen2024combining}. Say, Do, Understand specifies how to capture, prepare, analyze, and combine three evidence streams. Our application combines evidence at the participant level. It compares communication, activity, and explanation measures without claiming that a particular prompt produced a later action or understanding score. This keeps the value of combined evidence while keeping interpretation within the analysis we performed.

The methodological challenge is to keep each source's contribution visible after combining them. A profile should let readers see whether a communication pattern accompanies an activity pattern, while the explanation score remains visible as separate evidence. This structure also makes mismatches informative. A working implementation paired with a weak explanation need not collapse into a single judgment of effective use.

We also draw on LLM-assisted qualitative analysis~\cite{choudhuri2026copilot}. Our application uses independent model coding, rationales, researcher review, and repeated refinement of the codebook and rubrics. We report agreement for prompts and interview grades using Krippendorff's $\alpha$. These estimates describe agreement within the analyzed sample. The workflow complements interaction frameworks and assessment approaches by making their evidence, units, and limits of interpretation explicit.

\section{The Say, Do, Understand Workflow}
\label{sec:workflow}

\begin{figure*}[t]\centering\includegraphics[width=\textwidth]{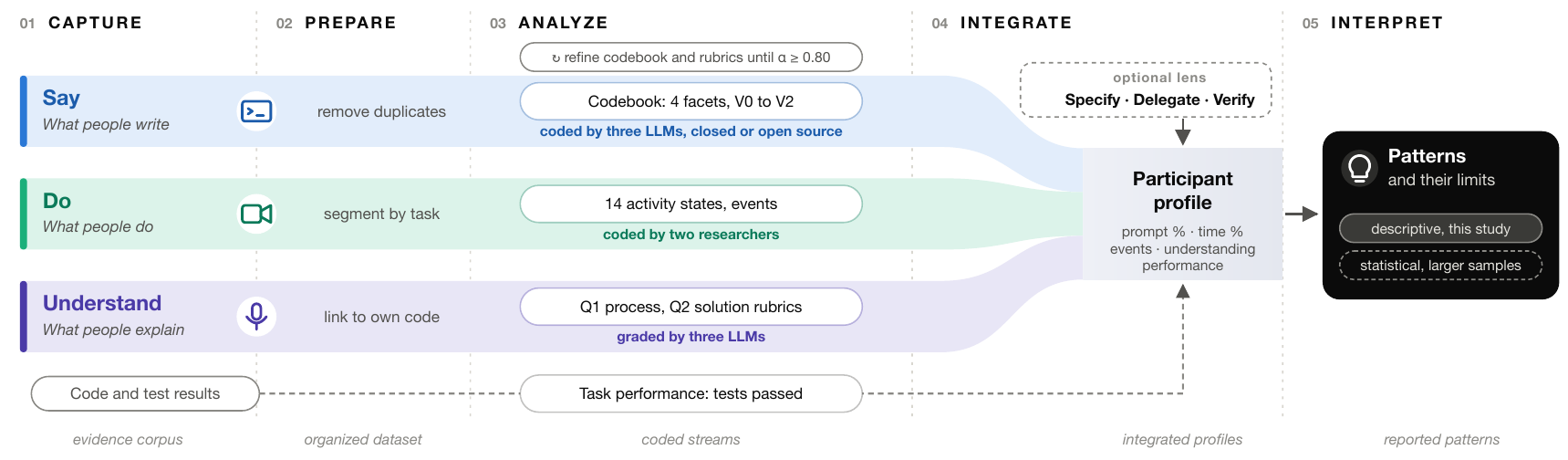}\caption{Say, Do, Understand. Capture and Prepare build the evidence corpus. Analyze processes prompts, screen-recorded behavior, and post-task explanations in parallel, alongside task performance from automated tests. Integrate builds participant-level profiles. Interpret reports patterns and their limits, descriptively in this study or statistically with larger samples. Our application uses Specify/Delegate/Verify as an optional lens during integration.}\label{fig:workflow}\end{figure*}

\subsection{Three Evidence Streams}

Figure~\ref{fig:workflow} shows the workflow: three evidence streams pass through five stages.

\textit{Say} is what a person writes to the AI. Its unit is a submitted prompt. Analysis can describe what the prompt refers to, what context it supplies, and how it directs the agent. Say records a request; it is not evidence that either party carried it out.

\textit{Do} is the visible activity around the interaction. Its units are sustained activity states and discrete events. States describe how participants spend their time. Events describe actions such as starting another chat. Do captures observable behavior, not the participant's purpose or understanding.

\textit{Understand} is what people explain after working on a task. We separate process understanding, the account of approach and progress, from solution understanding, the account of the implemented change. Its units are answers graded with separate rubrics. We grade solution explanations against the participant's committed code.

Each stream supports a different claim. A verification request belongs to Say. A developer-run test belongs to Do. An explanation of the solution belongs to Understand. The workflow connects these observations while keeping their original meaning.

\subsection{Five Stages and Their Artifacts}

\textit{Capture} decides what evidence to collect and which tasks it describes. In our application, this includes prompts, screen recordings, committed code, test outcomes, and interviews. The reusable artifact is an evidence corpus that links communication and activity to the implementation and its explanation. Task outcomes accompany the streams as a separate reference.

\textit{Prepare} makes the corpus ready for analysis. In our application, this meant removing duplicate or irrelevant prompts, segmenting recordings by task, and linking each explanation to the participant's own code. The reusable artifact is a cleaned, organized dataset with explicit inclusion decisions. Preparation should separate completed tasks from work in progress. Both can support an explanation, but they do not imply the same performance.

\textit{Analyze} processes the three streams in parallel. Prompt analysis applies a codebook. Video analysis identifies states and events. Explanation assessment applies the process and solution rubrics. Task performance comes from automated test results. Each instrument needs explicit category boundaries or grading criteria. In our application, three LLMs coded prompts and graded explanations, two researchers coded the video, and researchers refined the codebook and rubrics until $\alpha \geq 0.80$ (Section~\ref{sec:instruments}). The reusable artifacts are coded prompts, behavioral records, and graded explanations, together with the definitions used to produce them.

\textit{Integrate} combines these outputs into participant-level profiles. Each profile keeps prompt proportions, activity percentages, event counts, understanding scores, and task outcomes. The reusable artifact is a set of integrated profiles that lets readers inspect these measures together without reducing them to a single score. A conceptual lens may organize this comparison if its mapping stays explicit.

\textit{Interpret} identifies patterns that the profiles support and states what they leave open. The reusable artifact is a report of patterns that names the measures behind each claim and its scope. Our application describes participant-level patterns. It does not reconstruct the order of exchanges or identify causes. Studies with larger samples could use statistical analysis in this stage to test whether the patterns hold, for example associations between development behaviors and performance or understanding.

\subsection{Optional Lens and Analysis Principles}

We use Specify/Delegate/Verify only within Integrate. Specify groups prompt referents and supplied context. Delegate groups instructions, constraints, session-management events, and requests for agent checks. Verify groups developers' own checking activities. This lens guides how we read the application. The three streams and five stages do not depend on it.

Three principles guide that reading. First, keep requests for agent verification separate from the person's own checks. Asking for a test and running a test are different evidence. Second, report each checking activity separately instead of adding them into one total. Command-line testing, application inspection, developer-tool use, and output review stay distinct measures. Third, report activities with an unclear purpose as time use only. Viewing code can serve several purposes, so its duration alone should not decide what it means.

These definitions and stage outputs answer RQ1. They specify an end-to-end analysis whose intermediate artifacts can be inspected and adapted. Whether the instruments transfer to other populations or tools needs further evaluation.

\section{Analysis Instruments}
\label{sec:instruments}

\subsection{Say: Coding Prompts}

The Say instrument describes the content and function of developers' requests. We analyzed prompts with a human-in-the-loop procedure adapted from Choudhuri et al.~\cite{choudhuri2026copilot}. The procedure has four steps: discovering candidate codes, reconciling them, coding systematically, and refining. It produces prompt-level codes that can be summarized for each participant. Agent responses help us understand some prompts, but we do not code the agent as a communication partner.

\begin{figure*}[t]
    \centering
    \includegraphics[width=\textwidth]{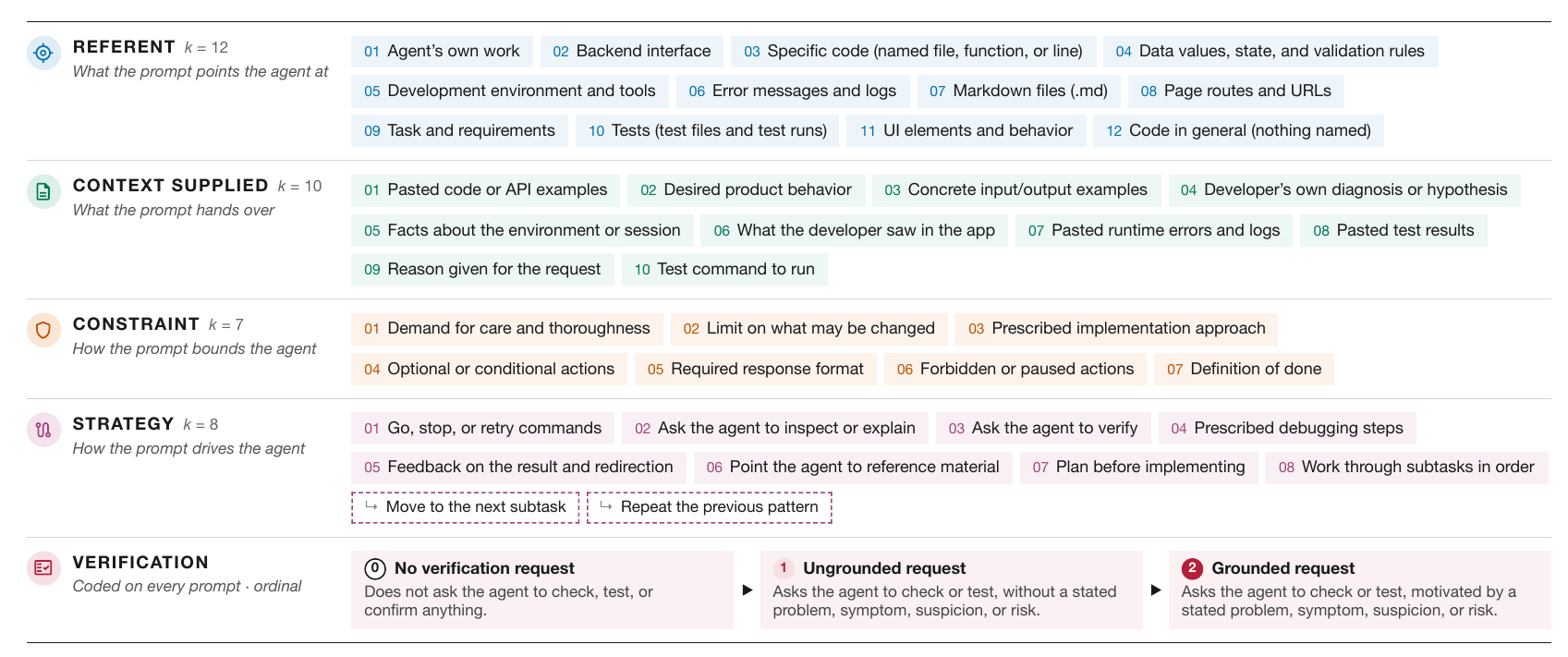}
    \caption{Prompt coding framework. We coded each prompt on four facets (referent, context supplied, constraint, and strategy) and separately classified its verification request as none, ungrounded, or grounded.}
    \label{fig:prompt_coding}
\end{figure*}

After we removed duplicates, system-generated notifications, and three off-task prompts, 380 prompts remained. Three models independently proposed themes for referent, context supplied, constraint, and strategy: Claude Opus-5, GPT-6-Astra, and Gemini-3.6-Flash. No model saw another model's proposals. A researcher merged the proposals into a common codebook before systematic coding. This gave shared definitions built from themes found in separate analyses.

Figure~\ref{fig:prompt_coding} summarizes the framework. The four facets answer different questions about a request. Referent is what the developer asks the agent to address. Context supplied is the information offered to help the agent act. Constraint captures limits or completion requirements. Strategy describes how the developer steers the exchange. For example, a request can refer to interface behavior, include runtime evidence, or redirect work after a result. The facets keep these distinctions instead of reducing a prompt to whether it asks for code.

Each model then coded every prompt using only the merged codebook. Each model saw the prompts in its own random order. Models gave a rationale before assigning codes and flagged data-quality problems instead of forcing them into categories. We assigned a code when at least two models agreed. Researchers used the rationales to examine disputed boundaries, because agreement alone does not explain why a code applies.

We measured agreement with Krippendorff's $\alpha$. Researchers examined 179 disputed prompt--code pairs for the seven codes with the lowest agreement. They refined the definitions, and fresh model sessions recoded all prompts. The final codebook has 37 codes (12 referents, 10 context codes, 7 constraints, and 8 strategies) plus 2 subcodes. Every code reached $\alpha \geq 0.80$. Mean $\alpha$ was 0.903, ranging from 0.81 to 1.00.

We also coded whether each prompt asked the agent to check its work. To judge this, we read each prompt together with the agent response just before it. V0 means the prompt did not ask for a check. V1 means it asked for a check without giving a reason. V2 means it asked for a check because of a specific problem, symptom, suspicion, or risk. Agreement was $\alpha=0.867$. We used the agent's response only to understand the prompt; we did not code the response itself.

\subsection{Do: Coding States and Events}

The Do instrument separates time-consuming activities from actions that happen at a single moment. We adapted the behavioral scheme used by Shihab et al.~\cite{shihab2025effects}. Two researchers independently coded a subset of recordings, discussed disagreements, and refined the codebook by adding emergent codes. After finalizing it, both independently coded 20\% of the video data, reaching Krippendorff's $\alpha$ = 0.80. They split the remaining recordings evenly.

\begin{table*}[t]
\centering
\small
\caption{Behavioral video codebook. States are mutually exclusive and measured as a percentage of time; events are discrete actions measured as counts.}
\label{tab:video-codebook}

\setlength{\tabcolsep}{5pt}
\renewcommand{\arraystretch}{1.12}

\begin{tabularx}{\textwidth}{@{}l X @{\hspace{1.5em}} l X@{}}
\toprule
\multicolumn{2}{c}{\textbf{States}} &
\multicolumn{2}{c}{\textbf{Events}} \\
\cmidrule(r){1-2}\cmidrule(l){3-4}
\textbf{Behavior} & \textbf{Definition} &
\textbf{Behavior} & \textbf{Definition} \\
\midrule

\rowcolor{gray!10}
View Task &
Read task instructions or API documentation. &
Approve Agent Action &
Approve an agent-requested command or tool action. \\

View Code &
Browse, search, or read source code. &
Reject Agent Action &
Reject an agent-requested command or tool action. \\

\rowcolor{gray!10}
View Web &
Consult external web resources. &
Reject Diff &
Reject or revert a proposed change. \\

View App &
Inspect the running application or database. &
Undo Agent Step &
Roll back an action performed by the agent. \\

\rowcolor{gray!10}
View Dev Tools &
Inspect browser console, network, or elements. &
Switch Copilot Mode &
Switch among Ask, Edit, Agent, or Plan modes. \\

Write Prompt &
Compose or paste a prompt to Copilot. &
Start New Chat &
Create a new Copilot chat session after the initial session. \\

\rowcolor{gray!10}
View Copilot Response &
Read Copilot's response or edit preview. &
Stop AI Generation &
Interrupt an active agent generation. \\

Monitor Agent Execution &
Watch autonomous file, edit, or command execution. &
Use Custom Skill &
Invoke a custom skill through a slash command. \\

\rowcolor{gray!10}
Review Agent Output &
Inspect agent-generated changes before accepting or rejecting them. &
Monitor Token Usage &
Check token-usage information. \\

Write Code &
Manually write or modify source code. &
& \\

\rowcolor{gray!10}
Test via CLI &
Manually run or inspect command-line tests. &
& \\

Interact with Experimenter &
Talk with the experimenter. &
& \\

\rowcolor{gray!10}
Configure Agent &
Edit instruction, MCP, or other agent-configuration files. &
& \\

Idle &
No observable task-related activity. &
& \\

\bottomrule
\end{tabularx}
\end{table*}

Table~\ref{tab:video-codebook} defines fourteen states. They are mutually exclusive and cover all session time, so a participant is always in exactly one state. The states include viewing task materials, source code, web resources, the application, developer tools, or a Copilot response. Other states cover writing prompts, writing code by hand, configuring the agent, testing on the command line, talking with the experimenter, and being idle. Monitoring execution and reviewing agent output are separate states.

This separation matters when interpreting an interaction. Monitoring execution records visible attention while the agent reads files, edits, or runs commands. Reviewing output records inspection of generated changes before accepting or rejecting them. Viewing a response records reading the agent's reply or edit preview. These categories describe visible activity. None of them alone shows whether a participant understood the content or judged it correctly.

We summarize states as the percentage of time each participant spent in each activity. Events are instantaneous actions reported as counts. They include approving or rejecting an agent action, rejecting a diff, undoing a step, switching Copilot mode, starting another chat, stopping generation, invoking a custom skill, and checking token usage. Starting another chat does not count the first session. An event count is not a duration, so we never combine counts with time percentages.

Each participant record keeps states and events separate. Readers can then check whether a group mean reflects a common activity or just a few participants. The record also keeps actions that do not fit a chosen lens. Code viewing stays a time-use measure because the recording alone cannot show its purpose.

\subsection{Understand: Assessing Process and Solution Explanations}

The Understand instrument uses two post-task questions, graded with the rubrics in Table~\ref{tab:rubrics}. Q1 asks participants to explain their overall process with Copilot. Q2 asks them to explain the solution they implemented. Graders see each answer together with the task and the participant's committed code. Q1 grades the process account and does not score whether the implementation is correct. Q2 is graded against the participant's actual implementation, not an ideal solution or a general description of the feature.

\begin{table*}[t]
\centering
\small
\caption{Rubrics for process understanding (Q1) and solution understanding (Q2).}
\label{tab:rubrics}

\setlength{\tabcolsep}{5pt}
\renewcommand{\arraystretch}{1.15}

\begin{tabularx}{\textwidth}{@{}c l X @{\hspace{1em}} c l X@{}}
\toprule
\multicolumn{3}{c}{\textbf{Q1: Process Understanding}} &
\multicolumn{3}{c}{\textbf{Q2: Solution Understanding}} \\
\cmidrule(r){1-3}\cmidrule(l){4-6}

\textbf{Score} & \textbf{Level} & \textbf{Definition} &
\textbf{Score} & \textbf{Level} & \textbf{Definition} \\
\midrule

\rowcolor{gray!10}
0 &
No recoverable process &
No identifiable account of an action taken on the task. &
0 &
No demonstrated understanding &
No explanation of how the implemented solution works. \\

1 &
Actions only &
Identifies actions taken but does not explain their rationale or connect observations to decisions. &
1 &
Central model incorrect &
Misstates a central mechanism, such as where functionality resides, how data flows, or what triggers the behavior. \\

\rowcolor{gray!10}
2 &
Connected but incomplete &
Connects at least one task observation, action, or decision, but does not explain the process from approach to endpoint. &
2 &
Partial or mixed &
Contains an important error, leaves the central mechanism unexplained, or explains only peripheral parts of the change. \\

3 &
Complete and coherent &
Explains the initial approach, progression, decision rationale, and endpoint with clear connections among them. &
3 &
Correct but limited &
Explains the central mechanism correctly, but is narrow in scope or lacks mechanistic detail across the change. \\

\rowcolor{gray!10}
&
&
&
4 &
Correct, mechanistic, and broad &
Explains multiple mechanisms specifically and covers a substantial portion of the implemented change. \\

\bottomrule
\end{tabularx}
\end{table*}

Q1 ranges from 0 to 3. A score of 0 means the answer names no task action. A score of 1 names actions but does not connect observations to decisions or explain why. A score of 2 connects at least one observation, action, or decision but leaves the process incomplete. A score of 3 explains the approach, progress, rationale, and endpoint coherently. The rubric separates naming actions from explaining how the work developed.

Q2 ranges from 0 to 4. A score of 0 means no explanation of how the solution works. A score of 1 misstates a central mechanism. A score of 2 contains an important error, leaves out the central mechanism, or explains only minor changes. A score of 3 correctly explains the central mechanism but is limited in scope or detail. A score of 4 explains several mechanisms specifically and covers much of the implemented change. Correctness comes first, then mechanism, then breadth.

The same three models independently graded every answer and gave rationales. We measured agreement with ordinal Krippendorff's $\alpha$. When agreement fell below 0.80, researchers reviewed the disputed answers and rationales and refined the rubric. Fresh model sessions then regraded every answer. One refinement clarified that simply saying the solution works does not explain the endpoint of the process. Final agreement was $\alpha=0.88$ for Q1 and $\alpha=0.81$ for Q2. Each final grade was decided by a two-of-three vote.

These grades describe the understanding participants showed in their answers. They do not measure everything a participant may know. In the application, we summed Q1 and Q2 over Tasks~1 and 2, giving a score from 0 to 14. We left out later tasks because not all participants reached them.

\subsection{Integrate and Interpret}
\label{sec:integrate}

Integration combines each participant's prompt proportions, time percentages, event counts, test outcomes, and understanding scores. Group summaries average these participant-level values, so each participant counts equally. We keep performance and understanding separate so that we can examine how they align without assuming that a working implementation means an accurate explanation.

We grouped participants with two median splits, one for performance and one for understanding. High-performing participants (HP) passed more than 66 of 136 tests; the remaining participants formed the low-performance group (LP). The understanding median was 9.5. Scores of 10 or more defined high understanding (HU); lower scores defined low understanding (LU). Ties went to the low group. Each split puts five participants in each group. Crossing the two splits gives four quadrants, which we describe as personas in Section~\ref{sec:findings}.

We used Specify/Delegate/Verify as an optional organizing lens. Specify contains prompt referents and context. Delegate contains instructions, constraints, session-management events, and requests for agent verification. Verify contains developers' command-line testing, application inspection, developer-tool use, and review of agent output. Figure~\ref{fig:personas}b uses these identifiers (S, D, and VF).

For RQ2, we described each persona by the measures on which it stood out from the other three (Figure~\ref{fig:personas}b). When we compared the two splits directly, we reported the group means themselves. We checked each pattern against the participant records behind it and, where one participant drives a pattern, we name them. These are possible patterns worth further study, not claims of statistical significance. Interpretation stays at the participant level and does not claim order in time or causation.

\section{Worked Application}
\label{sec:application}

\subsection{Study Setting}

We applied the workflow in an IRB-approved exploratory laboratory study in March 2026. Ten experienced developers used GitHub Copilot in VS Code through AWS WorkSpaces on Zoom. All used Claude Sonnet 4.6. They worked on an unfamiliar application for up to two hours and 15 minutes. The application illustrates the workflow with a small set of participants, tasks, and tool conditions.

We recruited through LinkedIn, GitHub, Reddit, and Discord. Participants needed at least three years of professional MERN experience, daily use of GenAI coding assistants, experience with production web applications, and willingness to be recorded. We also screened each participant's LinkedIn and GitHub profiles to reassure their provided information. We refer to participants as P01--P10. All participants lived in the United States. Mean age was 32.20 years (SD = 3.26), and mean professional development experience was 6.50 years (SD = 2.99).

Participants worked on a closed-source MERN application for tracking golf rounds and competitions. No participant had seen it before. The codebase had roughly 67,000 lines of JavaScript, JSX, and CSS across 363 source files, not counting tests, configuration, and dependencies. Participants built a multi-tab tournament-creation wizard opened from the existing Competitions page. The backend endpoints were already implemented and documented on a Swagger page available throughout the session.

The six frontend tasks followed the wizard's dependency order. Task~1 opened the wizard, hid the application's top mode bar, and restricted access to the first tab until a tournament existed. Task~2 implemented the Basic Info tab, including names, dates, host details, administrators, uploads, validation, and saving. Task~3 covered the registration window, entry fees, processing-fee options, withdrawal deadline, capacity, and optional swag. Task~4 used color pickers to choose leaderboard and scorecard colors and saved them with the tournament. Task~5 specified the tournament's courses. Task~6 specified competitive divisions and their round dates. Task~1 was a simple interface operation. Task~2 was the core integration challenge because it was the first task that had to use the existing backend API endpoints. Through these endpoints, the new tab checked that the tournament's short name was unique, filled host fields from the current user, searched existing users to add admins, uploaded files, and saved the tab's data. The tasks (3-6) followed the same pattern of form, validation, and saving as in Task 2.

One written document described the required behavior for all tasks. Participants completed the tasks in order because later tabs relied on state set up earlier. They cannot move on to the next task until the current task passed all required tests. Automated Playwright suites contained 3, 35, 39, 35, 16, and 8 tests for Tasks~1--6, for a total of 136. Participants could run the tests at any time but could not see the test code. Before the formal study, we ran a sandbox pilot with three researchers in our team.

\begin{figure*}[t]\centering\includegraphics[width=\textwidth]{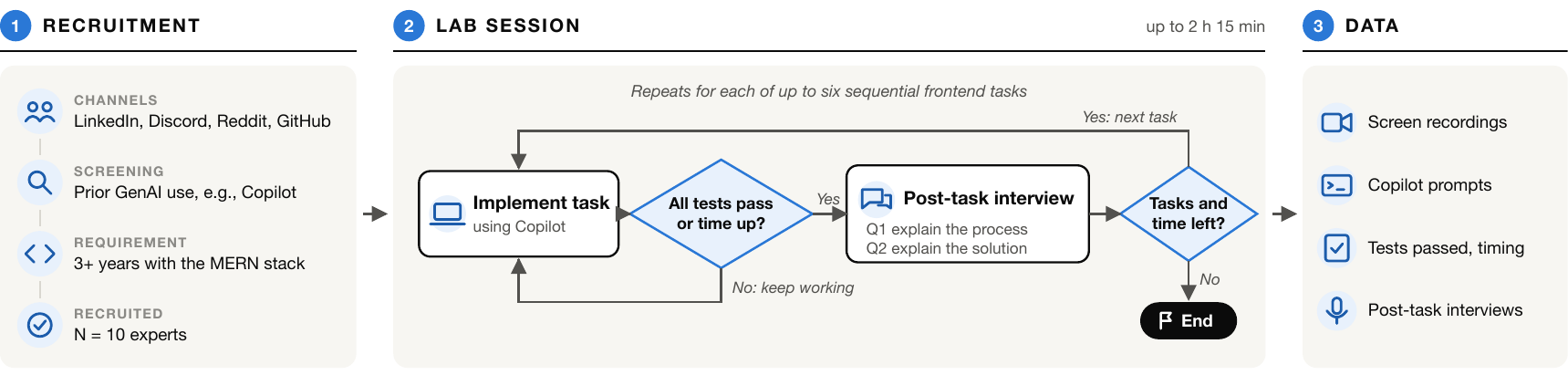}\caption{Study procedure. Ten experienced developers worked in order through up to six frontend tasks using GitHub Copilot, in sessions of up to two hours and 15 minutes. An interview followed each completed task, or the task in progress when the time limit was reached. We collected screen recordings, prompts, committed code, test outcomes and timing, and interviews.}\label{fig:procedure}\end{figure*}

Task completion time ran from when the participant said they were ready, after reading the instructions, until all required tests passed. Performance was the total number of tests passed within the time limit. An interview followed each completed task, or the task in progress when the time limit was reached. Participants explained their process and their implemented solution. The workflow therefore kept explanations of both finished and unfinished work, and solution grading used each participant's committed code.

We collected screen recordings, Copilot prompts, implementation and testing artifacts, and interviews. The instruments in Section~\ref{sec:instruments} produced participant-level profiles. Figure~\ref{fig:procedure} summarizes the procedure. The findings below answer RQ2. Activity and prompt summaries cover all tasks; understanding scores cover Tasks~1 and 2.

\begin{figure*}[t]
    \centering
    \includegraphics[width=\textwidth]{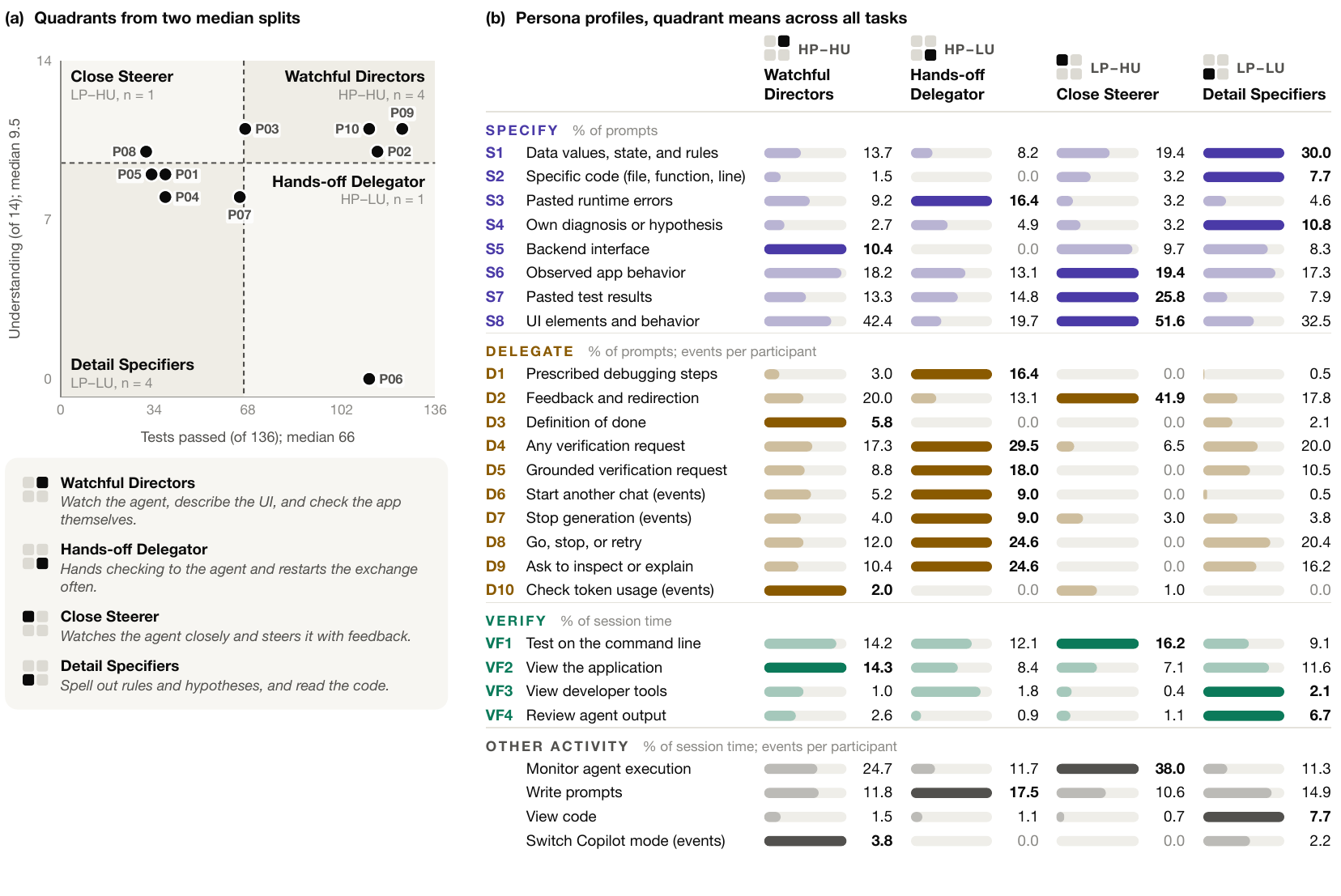}
    \caption{Quadrants and personas. (a) Each dot is one participant, labeled by ID. Dashed lines mark the medians: 66 of 136 tests passed and 9.5 of 14 understanding points (Tasks~1 and 2). Participants above a median form the high group; ties go to the low group. Crossing them gives four quadrants, each named for its persona. (b) Persona profiles as quadrant means across all tasks. In each row, bars scale to the highest of the four personas, and that value is bold. The glyph marks each persona's quadrant. Two quadrants hold one participant each and are cases, not types.}
    \label{fig:personas}
\end{figure*}

\subsection{Findings}
\label{sec:findings}

\subsubsection{Forming the Quadrants}
We formed four quadrants by crossing the two median splits described in Section~\ref{sec:integrate}. A participant was high performing (HP) if they passed more than 66 of the 136 tests and high understanding (HU) if they scored 10 or more of 14 on Q1 and Q2 for Tasks~1 and 2 (Table~\ref{tab:rubrics}). The two splits agreed for eight participants: four were HP--HU and four were LP--LU. Two participants crossed over. P06 passed 112 tests but scored 0 for understanding and said, ``No understanding; I didn't really look at any files.'' P08 passed only 31 tests but scored 10. Figure~\ref{fig:personas}a shows where each participant fell. Two placements sit close to the performance cut: P03 passed 67 tests and P07 passed 65.

For each quadrant, we built a persona from its profile across Say, Do, and Understand. Figure~\ref{fig:personas}b profiles each persona. The two crossover quadrants hold one participant each, so we treat them as cases, not types. The personas summarize this sample. They are patterns worth further study, not categories of developers.

\subsubsection{Four Personas}

\textbf{Watchful Directors (HP--HU, four participants)} passed 104.5 tests on average and explained their work well (10.8 of 14). \textit{Specify:} they described what they wanted in terms of the interface (S8, 42.4\% of prompts). \textit{Delegate:} they were the persona most likely to state a definition of done (D3), and two of them tracked token usage (D10). \textit{Verify:} they checked the running application most (VF2) and tested on the command line (VF1). \textit{Other activity:} they spent about a quarter of the session monitoring the agent (24.7\%), switched Copilot modes most often, and rarely opened the code (1.5\%).

\textbf{Hands-off Delegator (HP--LU, one participant)} passed 112 tests but could not explain the solution. \textit{Specify:} this participant pasted runtime errors more than any other persona (S3, 16.4\%). \textit{Delegate:} they handed the work and its checking to the agent: 29.5\% of prompts asked it to verify (D4), 24.6\% asked it to inspect or explain (D9), and 16.4\% prescribed debugging steps (D1). They also restarted often, with 9 new chats and 9 stopped generations. \textit{Verify:} they rarely reviewed agent output themselves (VF4). \textit{Other activity:} they spent the most time writing prompts (17.5\%), monitored the agent for 11.7\% of the session, rarely opened the code, and never switched modes. This case shows that passing tests and understanding the result can come apart.

\textbf{Close Steerer (LP--HU, one participant)} passed only 31 tests but explained the process and solution well. \textit{Specify:} their prompts described interface behavior (S8, 51.6\%) and pasted test results (S7, 25.8\%). \textit{Delegate:} they steered through feedback (D2, 41.9\% of prompts) and rarely asked the agent to verify (D4, 6.5\%). \textit{Verify:} they tested on the command line more than any other persona (VF1). \textit{Other activity:} they monitored the agent for 38.0\% of the session, more than any other persona, spent the least time writing prompts (10.6\%) and viewing code (0.7\%), and never switched modes. Their profile pairs long monitoring with slow progress.

\textbf{Detail Specifiers (LP--LU, four participants)} passed 43.5 tests on average and scored 8.5 for understanding. \textit{Specify:} their prompts spelled out implementation detail: data values, state, and rules (S1, 30.0\%), their own diagnoses (S4, 10.8\%), and specific code (S2, 7.7\%). \textit{Delegate:} they often issued short go, stop, or retry commands (D8, 20.4\%) and asked for verification in 20.0\% of prompts (D4). \textit{Verify:} they reviewed agent output more than any other persona (VF4). \textit{Other activity:} they read the most code (7.7\%), mostly one participant, monitored the agent least (11.3\%), and spent 14.9\% of their time writing prompts.

\subsubsection{Patterns Across Personas}
\textbf{In this sample, asking the agent to check did not go with checking oneself.} The two personas that asked most often for agent verification, the Hands-off Delegator and the Detail Specifiers, spent the least time testing on the command line. The two high-understanding personas asked for fewer checks and tested more themselves. Comparing the HU and LU groups, LU participants requested verification in 21.9\% of prompts and HU participants in 15.2\%, while their own checks differed by at most 4.8 points of session time.

\textbf{The two splits differ only through the crossover cases.} P06 is HP but LU, and P08 is LP but HU. As a result, the monitoring gap is 5.5 points between HP and LP but 16.0 points between HU and LU. The smaller gap rests on one participant: without P10, who monitored for 39.6\% of the session, it falls to 1.1 points. In the personas, the Close Steerer monitored for 38.0\% of the session and the Hands-off Delegator for 11.7\% (Figure~\ref{fig:personas}b). Likewise, feedback and redirection was close for HP and LP (18.6\% versus 22.6\% of prompts) but differed for HU and LU (24.4\% versus 16.8\%). Reading the personas alongside the group comparisons keeps these single-participant effects visible.

\section{Using the Workflow}
\label{sec:using}

Computing educators, industry practitioners, and researchers can adapt the workflow to their own settings. Each audience can keep Say, Do, and Understand separate while changing the tasks, collection procedures, and instruments. The uses below are prospective; our application shows one configuration, not effectiveness elsewhere.

\subsection{For Computing Educators}

Educators could start with a programming assignment, its grading criteria, and the code students submit. Capture would collect student prompts, screen activity, and explanations after the assignment. Prepare would link these records to each student's implementation and task status. The educator could then use the three instruments to examine what students asked for, what they did, and what they understood, alongside whether the code works.

The explanation rubrics are a starting point for separating how a student describes the process from how accurately they explain the solution. Adapting them would mean aligning solution grading with the assignment and each student's actual code. Process criteria would need to state how observations, actions, and decisions should connect. The current thresholds should not be assumed to suit learners without further evaluation.

The resulting profile could show where more evidence is needed. Correct code paired with a weak explanation could prompt an architectural walkthrough or an end-to-end trace of the change. A coherent explanation paired with incomplete functionality could prompt a look at the recorded work. These are possible teaching uses that we did not evaluate.

\subsection{For Industry Practitioners}

Industry practitioners could supply a development task, its acceptance criteria, and the resulting commit. Prompts and screen recordings would describe communication and activity around that work. Post-task explanations could cover the process and the central mechanisms of the implementation. The intended output is a profile that places task outcomes next to the evidence the developer gave the agent, the actions they took, and the change they can explain.

Adaptation would involve mapping the prompt and video codes to the team's tools and workflow. The current codebook and reliability estimates are therefore starting points for local refinement, not properties that carry over to a new setting.

Such an application could support the evaluation of code-review practices. Reviewers might ask a developer to explain a central mechanism, point to the relevant changes, describe a runtime check, or predict an edge case. The workflow would let them weigh these explanations separately from passing tests and from requests for agent checks. Whether this adds useful evidence in everyday development remains untested.

\subsection{For Researchers}

Researchers could reuse the five-stage structure while changing the programming tasks, participant population, or coding tool. They should document which codes and rubric boundaries they kept, changed, or removed. The main outputs would still be prompt codes, behavioral records, explanation scores, and combined profiles, each open to separate inspection. Another conceptual lens could replace Specify/Delegate/Verify without changing the three streams.

Several extensions follow from the application's limits. Larger and more varied samples could test whether the observed combinations recur. With enough participants, the same profiles could support statistical tests of associations between development behaviors and performance or understanding. New tasks could test whether the process and solution rubrics stay meaningful across implementations. Independent data are needed to evaluate coding and grading reliability beyond the refinement sample. Linking prompts and actions by timestamp, and coding agent responses as the partner's side of the exchange, would deepen the workflow.

\section{Limitations}
\label{sec:limitations}

We illustrate the workflow and instruments in a single study. Reuse in computing education and industry is a design goal, not an established result. The ten participants were US-based developers with MERN experience, using one codebase and GitHub Copilot with Claude Sonnet 4.6 in March 2026. The findings therefore cannot be generalized beyond this sample. Coding agents and their underlying models have already evolved since the study, so newer tools may change how developers specify, delegate, and verify. Eligibility was largely self-reported and included some inconsistencies, although we checked each participant's LinkedIn and GitHub profiles.

Screen recordings show visible interaction, not thinking. Window focus, cursor movement, and scrolling do not prove attention. Participants may read without interacting, or leave a pane open while looking elsewhere. Monitoring time may therefore overstate attention, and code-viewing time may understate reading. Eye tracking could show where participants look and so improve how the Do stream recognizes focus. Activity percentages also do not show how well participants checked their work. Being observed, fatigue, time pressure, or visible test failures may have shaped behavior during the timed session.

Understanding scores describe the explanations participants gave, not everything they knew. Process coherence and solution correctness are different constructs, even though we add them into one understanding score. Understanding covers Tasks~1 and 2, whereas performance and most activity summaries cover all tasks. We left later tasks out of the understanding score because not all participants reached them. These different scopes limit how directly the two measures can be compared.

Integration is at the participant level; it is not temporal or causal. The performance split (HP versus LP) and the understanding split (HU versus LU) each put five participants in each group, and the two splits differ only in where P06 and P08 fall. A pattern that appears under only one split therefore depends on these two participants. The personas summarize quadrant means, and two of them rest on a single participant. Median splits organize description; they do not establish population categories, statistical significance, or equivalence. Agent responses help interpret V0--V2, but we do not code the agent as a communication partner.

\section{Conclusion}
\label{sec:conclusion}

Say, Do, Understand is an end-to-end workflow for analyzing human--AI communication in programming. It moves prompts, screen recordings, and post-task explanations through five stages and analyzes them with three instruments: a 37-code prompt codebook, 14 activity states with nine events, and separate rubrics for process and solution understanding. Applied to ten experienced developers using GitHub Copilot, the workflow produced four personas. Performance and understanding agreed for eight developers but split for two: one passed 112 of 136 tests yet could not explain the solution, while another passed 31 and explained it well. The personas that most often asked the agent to verify its work spent the least time testing on the command line. Educators, practitioners, and researchers can adapt these instruments, and larger samples could test whether the patterns hold.

\bibliographystyle{IEEEtran}
\bibliography{bib}

\end{document}